\documentclass[fleqn,usenatbib]{mnras}
\usepackage{amsmath}	
\usepackage{amssymb}	

\usepackage{newtxtext,newtxmath}

\usepackage[T1]{fontenc}

\usepackage{graphicx}	

\newcommand \msun {M$_\odot$}

\newcommand \drvm  {$\Delta{\rm RV}_{\rm max}$}

\title[Limits on collisonal-triples]{Limits on a population of collisional-triples as progenitors of Type-Ia supernovae}

\author[N. Hallakoun \& D. Maoz]{
Na'ama Hallakoun\thanks{E-mail: \href{mailto:naama@wise.tau.ac.il}{naama@wise.tau.ac.il} (NH)}
and Dan Maoz
\\
School of Physics and Astronomy, Tel-Aviv University, Tel-Aviv 6997801, Israel\\
}

\date{Accepted XXX. Received YYY; in original form ZZZ}

\pubyear{2019}

\begin{document}
\label{firstpage}
\pagerange{\pageref{firstpage}--\pageref{lastpage}}\maketitle
\begin{abstract}
The progenitor systems of Type-Ia supernovae (SNe Ia) are yet unknown. The collisional-triple SN Ia progenitor model posits that SNe Ia result from head-on collisions of binary white dwarfs (WDs), driven by dynamical perturbations by the tertiary stars in mild-hierarchical triple systems. To reproduce the Galactic SN Ia rate, at least $\sim 30-55$~per cent of all WDs would need to be in triple systems of a specific architecture. We test this scenario by searching the \textit{Gaia} DR2 database for the postulated progenitor triples. Within a volume out to 120\,pc, we search around \textit{Gaia}-resolved double WDs with projected separations up to 300\,au, for physical tertiary companions at projected separations out to 9000\,au. At 120\,pc, \textit{Gaia} can detect faint low-mass tertiaries down to the bottom of the main sequence and to the coolest WDs. Around 27 double WDs, we identify zero tertiaries at such separations, setting a 95~per cent confidence upper limit of 11~per cent on the fraction of binary WDs that are part of mild hierarchical triples of the kind required by the model. As only a fraction (likely $\sim 10$~per cent) of all WDs are in $<300$\,au WD binaries, the potential collisional-triple progenitor population appears to be at least an order of magnitude (and likely several) smaller than required by the model.
\end{abstract}



\section{Introduction}
\label{sec:Intro}

A major unsolved puzzle in astrophysics is the identity of the progenitors of Type-Ia supernovae (SNe~Ia; see \citealt{Maoz_2014, Livio_2018, Wang_2018}, for a review). A number of progenitor scenarios have been considered over the years. Among them, several have involved the collision of two white dwarfs (WDs) in a variety of configurations and environments \citep[e.g.][]{Benz_1989, Thompson_2011}. In particular, \citet{Katz_2012} and \citet{Kushnir_2013} have proposed that SNe~Ia result from head-on collisions of WDs in ``mild-hierarchical'' triple systems. The triple system consists of an inner double-WD binary with separation $a\lesssim 300$\,au, orbited by a roughly solar-mass tertiary star in an orbit with pericentre separation $\sim 3-10$ times the inner-binary's separation. Using numerical integration of the evolution of such 3-body systems and assuming a uniform distribution of inclinations, \citet{Katz_2012} found that, in about 5~per cent of all such systems (the 5~per cent within a small range around a high inclination between the inner and outer orbits), the outer tertiary stochastically drives a Kozai-Lidov perturbation of the inner pair's orbital eccentricity. Within a few Gyr, the eccentricity lands on a high-enough value to send the inner pair on a head-collision. Given that the actual distribution of inclinations of the systems that survive as a wide double WD with a tertiary companion is unlikely to be uniform, this 5~per cent is most likely an upper limit of the fraction of systems that undergo a collision. \citet{Kushnir_2013} showed that the compression undergone by the WDs upon collision could be effective both at igniting a thermonuclear carbon detonation, and in producing about 0.5\,\msun\ of radioactive $^{56}$Ni in the explosion, as observed in typical SN Ia events. The range of masses of the WDs that collide could also reproduce the observed range of SN Ia luminosities and their correlation with light-curve evolution time. The asymmetry of the system at the time of explosion predicts double-peaked emission line profiles in SN Ia spectra during the nebular phase in a fraction of events, for which there may be some observational evidence \citep{Dong_2015, Dong_2018, Kollmeier_2019, Vallely_2019}.

However, a major challenge for the collisional-triple model is to produce a collision rate that can match the SN Ia rate in our Galaxy. The Milky Way, if it is a typical Sbc galaxy, has a SN Ia rate per unit stellar mass of $(1.12\pm0.35)\times 10^{-13}$~yr$^{-1}$\,\msun$^{-1}$ \citep{Li_2011, Graur_2017}.\footnote{For a total stellar mass of  $(6.4\pm0.6)\times 10^{10}$\,\msun\ \citep{Mcmillan_2011}, the Galactic rate is $(7.2\pm2.3)\times 10^{-3}$\,yr$^{-1}$, or a SN~Ia every $100-200$\,yr. This is broadly consistent with the four known Galactic events over the last millennium that were likely or certain SNe~Ia (SN1006, SN1572-Tycho, SN1604-Kepler, and G1.9+03).} The stellar mass density in the Solar neighborhood is $0.085\pm 0.010$\,\msun\,pc$^{-3}$ \citep{Mcmillan_2011}, and the WD number density is $0.0045\pm 0.0004$~pc$^{-3}$ \citep{Hollands_2018}, giving a stellar-mass to WD number ratio of $18.9\pm 2.8$\,\msun\,WD$^{-1}$. The SN~Ia rate per WD is therefore $(2.1\pm0.7)\times 10^{-12}$\,yr$^{-1}$, and the fraction of all WDs that explode as SNe Ia during the 10~Gyr-lifetime of the Galaxy is $(2.1\pm 0.7)$~per cent. Considering then, that in the collisional-triple model at most 5~per cent of suitable triple systems undergo a SN~Ia inducing collision, implies that at least $\sim 30-55$~per cent of all WDs need to be in suitable triple systems, if this mechanism is to explain all SNe~Ia \citep[see also][]{Papish_2016}.

While the observational picture on stellar multiplicity is far from clear yet, and even more so the situation regarding the multiplicity of stellar remnants such as WDs, there are already question marks as to the plausibility of such a high frequency of triples with a specific architecture [$a<300$\,au, $r_{\rm peri, out}=(3-10)a$]. The fraction of triple stars (of all configurations) among main-sequence A-through-K type stars has been estimated at $5-15$~per cent \citep[][and references therein]{Tokovinin_2008, Duchene_2013, Tokovinin_2014, Leigh_2013}, or $15-30$~per cent \citep{Moe_2017}, and likely only a fraction of those have the required architectures. \citet{Klein_2017} estimated the occurrence frequency of  the inner binary WD component of the collisional-triple model, by analysing two published studies of $>1 M_\odot$ stars (that will eventually evolve into WDs): an adaptive optics survey of A stars by \citet{DeRosa_2014}, in which $>1$\,\msun\ companions within 400\,au were searched for; and a radial-velocity survey by \citet{Mermilliod_2007} of $>1$\,\msun\ giants, which was searched for $>1$\,\msun\ companions within 3\,au. \citet{Klein_2017} conclude that $15-20$~per cent of the intermediate mass stars that become WDs are in binaries that can constitute the inner component of collisional-triple systems, and therefore even if all such binaries had a mild hierarchical tertiary, there would still be a factor 2 shortage of collisional-triple progenitors for SNe Ia. Even if triple main-sequence systems are abundant, a challenge of the model that was already acknowledged by \citet{Katz_2012} is that triple systems with the suitable architecture and relative orbital inclination to induce to a collision of the inner binary will undergo such a collision already when the stars are on the main sequence (without an ensuing SN~Ia), effectively eliminating all the SN Ia progenitor systems. This conclusion has emerged also from binary population synthesis calculations \citep{Hamers_2013, Toonen_2018}. \citet{Katz_2012} raised the possibility that angular momentum loss by the system due to asymmetric mass loss during the evolution to the WD stage, or perturbations by passing stars, could ``reset'' the relative orbital inclinations of the system, and thus solve the problem.

In this paper, we address more directly the subject of the putative triple-progenitor population of SNe Ia, by searching the \textit{Gaia} DR2 database \citep{Gaia_2016, GaiaDR2_2018} specifically for double WDs orbited by a tertiary star, as envisaged in the collisional-triple model.
 
\section{\textit{Gaia} search}
\label{sec:Gaia}

We search for triple systems akin to those required by the collisional-triple model using two approaches: first, by identifying resolved WD binaries with projected separations $<300$\,au, and searching their surroundings for tertiaries with projected separations $<9000$\,au; and, second, by searching for tertiaries projected within $3000$\,au of unresolved, double-WD,  separation $a\sim 0.1-1$\,au, candidates identified via radial-velocity variations. Our search for triples is conservatively inclusive in several respects. First, by pre-selecting WD binaries, and then asking what fraction of those binaries have a tertiary, we obtain an upper limit on the fraction of \textit{all} WDs in triples, since not all WDs are in binaries. Second, because of projection effects, some of the selected inner binaries will have physical separations $a>300$\,au, and some of the counted tertiaries will be at physical separations beyond the tertiary separation range required by the model. By including all of these systems, we will obtain a conservative upper limit on the true fraction of of WDs that are in triples with the architecture required by the model.

\subsection{Resolved \textit{Gaia} triples}
Following \citet{GentileFusillo_2019}, we start by identifying all WD candidates using an initial colour-magnitude cut in \textit{Gaia} DR2 data:
\begin{align}
\texttt{parallax\_over\_error} &> 1\\
M_G &> 5\\
M_G &> 5.93 + 5.047\times \left(G_\textrm{BP}-G_\textrm{RP}\right)\\
M_G &> 6 \times \left(G_\textrm{BP}-G_\textrm{RP}\right)^3+\nonumber\\
 &- 21.77\times \left(G_\textrm{BP}-G_\textrm{RP}\right)^2 +\nonumber\\
&+27.91\times \left(G_\textrm{BP}-G_\textrm{RP}\right)+0.897\\
\left(G_\textrm{BP}-G_\textrm{RP}\right) &< 1.7
\end{align}
where $M_G = \texttt{phot\_g\_mean\_mag}+5\times\log_{10}\left(\texttt{parallax}\right)-10$ is the absolute magnitude in the $G$ band. This results in $8,144,732$ sources.
We further select only those within a distance of 120\,pc:
\begin{equation}
\texttt{parallax} > 8.33
\end{equation}
leaving $313,871$ sources.
We follow \citet{GaiaHRD_2018} and remove astrometric artefacts by requiring:
\begin{equation}\label{eq:astrometric}
\sqrt{\frac{\chi^2}{\nu'-5}}<1.2 \textrm{max}\left(1,~ e^{-0.2\left(G-19.5\right)}\right)
\end{equation}
where $\chi^2$ is \texttt{astrometric\_chi2\_al} and $\nu'$ is \texttt{astrometric\_n\_good\_obs\_al} in the \textit{Gaia} database, leaving $86,182$ sources.

Since the $G_\textrm{BP}$ and $G_\textrm{RP}$ fluxes are calculated by integrating over low-resolution spectra, they are more prone to contamination from nearby sources, compared to the $G$-band flux, which is measured by photometric profile fitting. Following \citet{ElBadry_2018} and \citet{Evans_2018}, we filter out sources that might be contaminated by nearby sources by limiting the total $G_\textrm{BP}$ and $G_\textrm{RP}$ excess compared to the $G$ band:
\begin{align}
\texttt{phot\_bp\_rp\_excess\_factor} &> 1.0 + 0.015 \left(G_\textrm{BP}-G_\textrm{RP}\right)^2\\
\texttt{phot\_bp\_rp\_excess\_factor} &< 1.3 + 0.06 \left(G_\textrm{BP}-G_\textrm{RP}\right)^2
\end{align}
leaving $26,591$ sources.

We further follow \citet{ElBadry_2018} by selecting only sources with high signal-to-noise ratio photometry, i.e., $<2$~per cent flux uncertainty in the $G$ band, and $<5$~per cent in both the $G_\textrm{BP}$ and $G_\textrm{RP}$ bands:
\begin{align}
\texttt{phot\_g\_mean\_flux\_over\_error} &> 50 \\
\texttt{phot\_rp\_mean\_flux\_over\_error} &> 20 \\
\texttt{phot\_bp\_mean\_flux\_over\_error} &> 20
\end{align}
leaving $17,410$ sources. These criteria assure that we only select sources that fall with high certainty within in the WD region of the colour-magnitude diagram.

Finally, we select only sources where the relative error on the parallax is smaller than $10$~per cent:
\begin{equation}
\texttt{parallax\_over\_error}>10
\end{equation}
resulting in $17,395$ sources.

We then choose all physical-double resolved WDs within 120\,pc with projected separations $<300$\,au, by requiring
\begin{equation}
\frac{\theta}{\textrm{arcsec}} \leq 10^{-3}\frac{s}{\textrm{au}}\frac{\varpi}{\textrm{mas}}
\end{equation}
where $\theta$ is the angular separation, $s=300$\,au is the projected separation, and $\varpi$ is the parallax.
Following \citet[][eq. 2]{ElBadry_2018} we further select only pairs with consistent parallaxes:
\begin{equation}\label{eq:distance}
\Delta d - 2s \leq 3 \sigma_{\Delta d},
\end{equation}
where $\Delta d = \left| 1/\varpi_1 - 1/\varpi_2 \right|$~pc, $s = \theta /\varpi_1$~pc, and $\sigma_{\Delta d}= \sqrt{\sigma_{\varpi,1}^2/\varpi_1^4 + \sigma_{\varpi,2}^2/\varpi_2^4}$~pc. Here $\varpi_i$ and $\sigma_{\varpi,i}$ are the parallax of the $i$th target and its standard error in arcsec, respectively, and the angular separation, $\theta$, is in radians. This means that we inclusively count a system as a WD binary, even if its component distances differ at the $3 \sigma_{\Delta d}$ level.

\begin{figure}
\includegraphics[width=\columnwidth]{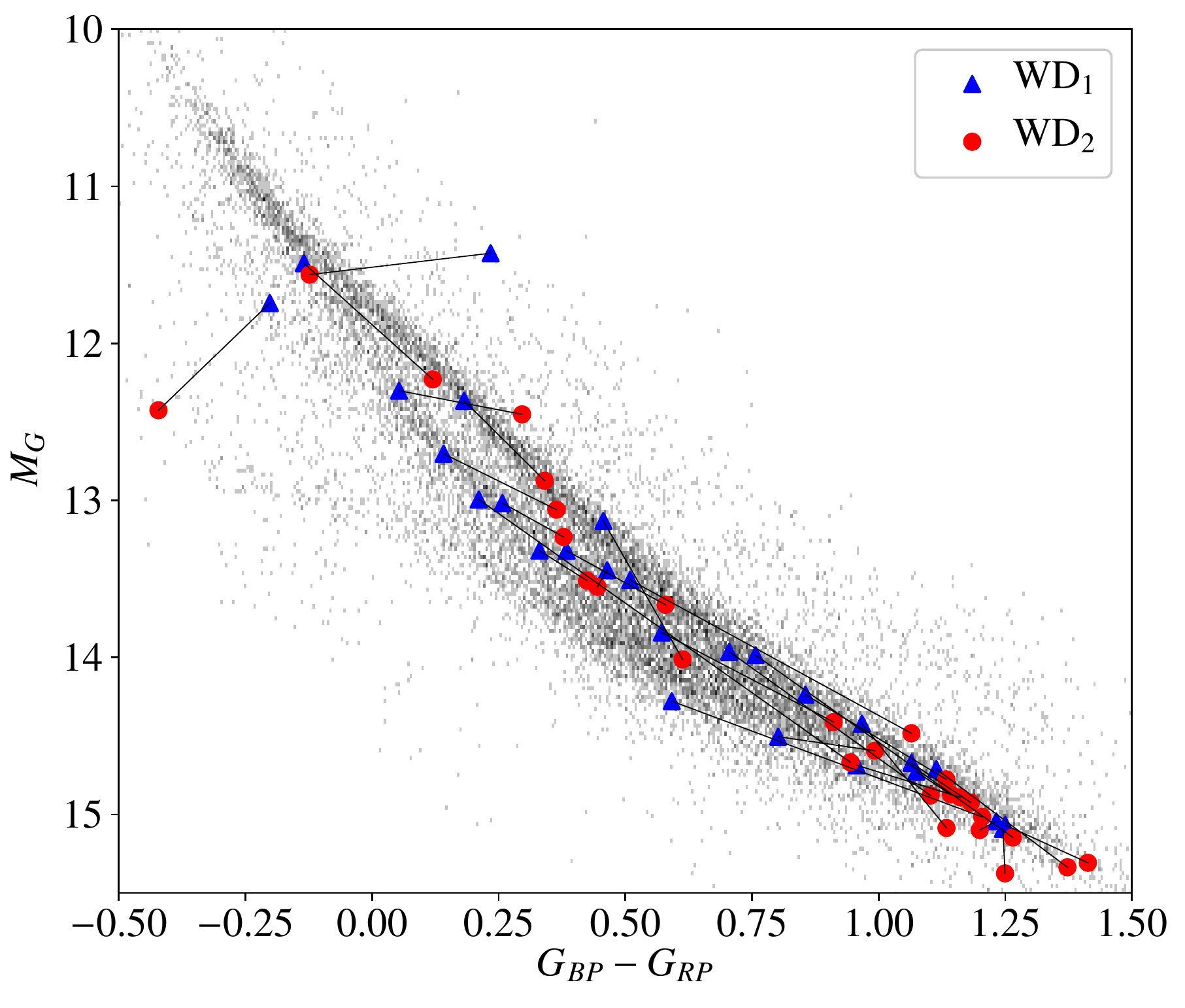}
\caption{\textit{Gaia} colour-magnitude diagram for resolved double WDs within 120\,pc, with projected separations $s<300$\,au. Pairs are connected by black solid lines, where the photometric primary (secondary) is marked by a blue triangle (red circle). None of the pairs have a tertiary companion with a projected separation $s<9000$\,au. The number density in this parameter space of the full 120\,pc WD sample from \textit{Gaia} ($17,395$ sources) is shown in grayscale for reference.}
\label{fig:Resolved}
\end{figure}

The search results in $27$ WD pairs, listed in Table~\ref{tab:DWD}, out of the $17,395$ WDs that are nearer than 120\,pc. The relative projected velocity differences between the members of each pair, also listed in Table~\ref{tab:DWD}, are well below the $\sim 6$\,km\,s$^{-1}$ maximum that is possible between bound solar-mass objects, indicating these are real, bound, pairs. Fig.~\ref{fig:Resolved} shows the WD-pair locations on the \textit{Gaia} colour-magnitude diagram. 

We note, following \citet{ElBadry_2018} and \citet{Arenou_2018}, that the above 27 resolved pairs are a small fraction of the actual number of WD pairs with separations $<300$\,au, which is likely to be at least 10~per cent \citep{Maoz_2018}. First, \textit{Gaia} is incomplete to binaries at angular separations $\theta\lesssim 0.7$\,\arcsec, corresponding to 84\,au at 120\,pc. More importantly, the \textit{Gaia} $G_\textrm{BP}-G_\textrm{RP}$ colour is not available, in most cases, for both components of doubles with projected separations $\theta<2$\,\arcsec\ (240\,au at 120\,pc), selecting against the identification of one or both stars as WDs. We further note that extending the WD sample to 200 or 300\,pc did not result in any additional $<300$\,au WD pairs, likely because of the limitations imposed by the resolution, which become more severe with distance. However, our strategy for testing the collisional-triple model does not require a complete census of WD binaries; to the contrary, we select only a small minority of double-WDs that \textit{Gaia} can detect, but then perform a thorough search for tertiaries around these binaries.  

By limiting our sample of WD pairs to 120\,pc, we expect that, at the limiting \textit{Gaia} magnitude $G \sim 20$, we are sensitive and largely complete to tertiary companions corresponding to the faintest, lowest-mass, main-sequence stars, as well as the coolest WDs. As shown by \citet[][fig. 19]{GentileFusillo_2019}, the 100\,pc \textit{Gaia} DR2 WD sample is complete up to $M_G \sim 16$, indeed corresponding to all but the coolest few percent of WDs with $M_G > 16$ \citep{Isern_2019, Tremblay_2019}. This completeness likely applies also at the relatively brighter luminosities near the bottom of the main sequence ($M_G \sim 14$, $T_\textrm{eff} \lesssim 3000$\,K), corresponding to $\lesssim 0.2$\,\msun\ M-dwarfs \citep{Hillenbrand_2004}. However, a more thorough future assessment of the main-sequence completeness at 120\,pc will place our conclusions, below, on a surer footing.

\begin{table*}
\caption{Resolved double WDs within 120\,pc, with projected separations $s<300$\,au. Each system appears as a pair of rows separated by a line. Columns give \textit{Gaia} DR2 ID, parallax, distance, and proper motion. The third row of each system lists its angular and projected separations, and the relative proper motion. None of these WD pairs have a tertiary star within a projected separation of 9000\,au.}
\label{tab:DWD}
\begin{center}
\begin{tabular}{c c c c c c c c}
\hline
GaiaID & $\varpi$ (mas) & $d$ (pc) & $\mu_\textrm{RA}$ (mas\,yr$^{-1}$) & $\mu_\textrm{Dec}$ (mas\,yr$^{-1}$) & $v_\textrm{RA}$ (km\,s$^{-1}$) & $v_\textrm{Dec}$ (km\,s$^{-1}$)\\
\hline
4617812054836337536 & $17.83 \pm 0.19$ & $56.07 \pm 0.60$ & $58.55 \pm 0.41$ & $-190.23 \pm 0.33$ & $15.56 \pm 0.20$ & $-50.56 \pm 0.55$ \\
4617812054837799168 & $18.12 \pm 0.17$ & $55.19 \pm 0.52$ & $57.93 \pm 0.36$ & $-191.60 \pm 0.29$ & $15.16 \pm 0.17$ & $-50.13 \pm 0.48$ \\
 & \multicolumn{2}{c}{$\theta=3.92450 \pm 0.00011$\,\arcsec} & \multicolumn{2}{c}{$s=220.1 \pm 2.4$\,au} & \multicolumn{2}{c}{$v_\textrm{rel}=0.60 \pm 0.69$\,km\,s$^{-1}$}\\
\hline 
2482813425794003200 & $23.94 \pm 0.19$ & $41.77 \pm 0.33$ & $-264.84 \pm 0.38$ & $-265.05 \pm 0.17$ & $-52.45 \pm 0.43$ & $-52.49 \pm 0.42$ \\
2482813430089468160 & $23.36 \pm 0.19$ & $42.80 \pm 0.35$ & $-259.72 \pm 0.38$ & $-264.77 \pm 0.16$ & $-52.70 \pm 0.43$ & $-53.72 \pm 0.44$ \\
 & \multicolumn{2}{c}{$\theta=3.18446 \pm 0.00022$\,\arcsec} & \multicolumn{2}{c}{$s=133.0 \pm 1.1$\,au} & \multicolumn{2}{c}{$v_\textrm{rel}=1.26 \pm 0.71$\,km\,s$^{-1}$}\\
\hline 
4613612951211823616 & $33.859 \pm 0.036$ & $29.534 \pm 0.032$ & $-61.767 \pm 0.062$ & $-22.360 \pm 0.056$ & $-8.648 \pm 0.013$ & $-3.1306 \pm 0.0086$ \\
4613612951211823104 & $33.848 \pm 0.047$ & $29.543 \pm 0.041$ & $-73.700 \pm 0.078$ & $-11.815 \pm 0.072$ & $-10.322 \pm 0.018$ & $-1.655 \pm 0.010$ \\
 & \multicolumn{2}{c}{$\theta=7.127628 \pm 0.000039$\,\arcsec} & \multicolumn{2}{c}{$s=210.51 \pm 0.22$\,au} & \multicolumn{2}{c}{$v_\textrm{rel}=2.232 \pm 0.017$\,km\,s$^{-1}$}\\
\hline 
16166875378033664 & $10.16 \pm 0.19$ & $98.4 \pm 1.9$ & $55.98 \pm 0.32$ & $-60.44 \pm 0.27$ & $26.12 \pm 0.51$ & $-28.20 \pm 0.55$ \\
16166875377576064 & $10.59 \pm 0.35$ & $94.4 \pm 3.1$ & $56.62 \pm 0.57$ & $-58.12 \pm 0.48$ & $25.34 \pm 0.88$ & $-26.01 \pm 0.89$ \\
 & \multicolumn{2}{c}{$\theta=2.94598 \pm 0.00029$\,\arcsec} & \multicolumn{2}{c}{$s=290.0 \pm 5.5$\,au} & \multicolumn{2}{c}{$v_\textrm{rel}=2.3 \pm 1.3$\,km\,s$^{-1}$}\\
\hline 
3249479592235301376 & $32.16 \pm 0.11$ & $31.10 \pm 0.10$ & $-28.35 \pm 0.18$ & $-262.86 \pm 0.19$ & $-4.179 \pm 0.029$ & $-38.75 \pm 0.13$ \\
3249479592234269056 & $32.25 \pm 0.13$ & $31.01 \pm 0.13$ & $-8.52 \pm 0.22$ & $-264.95 \pm 0.25$ & $-1.252 \pm 0.033$ & $-38.95 \pm 0.16$ \\
 & \multicolumn{2}{c}{$\theta=5.40038 \pm 0.00012$\,\arcsec} & \multicolumn{2}{c}{$s=167.93 \pm 0.55$\,au} & \multicolumn{2}{c}{$v_\textrm{rel}=2.933 \pm 0.042$\,km\,s$^{-1}$}\\
\hline 
3404213863614488192 & $10.70 \pm 0.13$ & $93.5 \pm 1.2$ & $-18.61 \pm 0.21$ & $19.61 \pm 0.16$ & $-8.25 \pm 0.14$ & $8.69 \pm 0.13$ \\
3404213863611804672 & $10.73 \pm 0.12$ & $93.2 \pm 1.1$ & $-17.53 \pm 0.20$ & $17.31 \pm 0.15$ & $-7.74 \pm 0.12$ & $7.65 \pm 0.11$ \\
 & \multicolumn{2}{c}{$\theta=3.09811 \pm 0.00014$\,\arcsec} & \multicolumn{2}{c}{$s=289.6 \pm 3.6$\,au} & \multicolumn{2}{c}{$v_\textrm{rel}=1.16 \pm 0.21$\,km\,s$^{-1}$}\\
\hline 
3001062974507933952 & $16.85 \pm 0.25$ & $59.36 \pm 0.89$ & $-163.00 \pm 0.37$ & $-75.65 \pm 0.46$ & $-45.87 \pm 0.70$ & $-21.29 \pm 0.34$ \\
3001062974507934080 & $16.59 \pm 0.28$ & $60.3 \pm 1.0$ & $-168.99 \pm 0.42$ & $-75.74 \pm 0.50$ & $-48.29 \pm 0.83$ & $-21.65 \pm 0.39$ \\
 & \multicolumn{2}{c}{$\theta=4.35293 \pm 0.00033$\,\arcsec} & \multicolumn{2}{c}{$s=258.4 \pm 3.9$\,au} & \multicolumn{2}{c}{$v_\textrm{rel}=2.5 \pm 1.1$\,km\,s$^{-1}$}\\
\hline 
3144837318276010624 & $54.935 \pm 0.070$ & $18.203 \pm 0.023$ & $211.54 \pm 0.13$ & $-1782.658 \pm 0.077$ & $18.255 \pm 0.026$ & $-153.83 \pm 0.20$ \\
3144837112117580800 & $55.141 \pm 0.061$ & $18.135 \pm 0.020$ & $209.90 \pm 0.11$ & $-1790.484 \pm 0.063$ & $18.045 \pm 0.022$ & $-153.93 \pm 0.17$ \\
 & \multicolumn{2}{c}{$\theta=16.404756 \pm 0.000064$\,\arcsec} & \multicolumn{2}{c}{$s=298.62 \pm 0.38$\,au} & \multicolumn{2}{c}{$v_\textrm{rel}=0.231 \pm 0.082$\,km\,s$^{-1}$}\\
\hline 
1118460924702185088 & $21.83 \pm 0.15$ & $45.80 \pm 0.31$ & $-227.50 \pm 0.17$ & $-315.27 \pm 0.22$ & $-49.40 \pm 0.34$ & $-68.45 \pm 0.47$ \\
1118460924703171584 & $21.60 \pm 0.11$ & $46.29 \pm 0.24$ & $-224.95 \pm 0.12$ & $-311.65 \pm 0.17$ & $-49.36 \pm 0.26$ & $-68.39 \pm 0.36$ \\
 & \multicolumn{2}{c}{$\theta=3.53694 \pm 0.00012$\,\arcsec} & \multicolumn{2}{c}{$s=162.0 \pm 1.1$\,au} & \multicolumn{2}{c}{$v_\textrm{rel}=0.07 \pm 0.72$\,km\,s$^{-1}$}\\
\hline 
606938634805314944 & $17.817 \pm 0.077$ & $56.13 \pm 0.24$ & $-115.21 \pm 0.13$ & $-26.764 \pm 0.097$ & $-30.65 \pm 0.14$ & $-7.121 \pm 0.040$ \\
606938634805314816 & $17.747 \pm 0.066$ & $56.35 \pm 0.21$ & $-115.58 \pm 0.11$ & $-22.206 \pm 0.076$ & $-30.87 \pm 0.12$ & $-5.931 \pm 0.030$ \\
 & \multicolumn{2}{c}{$\theta=3.765706 \pm 0.000069$\,\arcsec} & \multicolumn{2}{c}{$s=211.36 \pm 0.91$\,au} & \multicolumn{2}{c}{$v_\textrm{rel}=1.210 \pm 0.034$\,km\,s$^{-1}$}\\
\hline 
798602271945568256 & $14.54 \pm 0.20$ & $68.79 \pm 0.96$ & $8.81 \pm 0.40$ & $-104.02 \pm 0.47$ & $2.87 \pm 0.14$ & $-33.92 \pm 0.50$ \\
798602267650286464 & $13.68 \pm 0.40$ & $73.1 \pm 2.1$ & $10.10 \pm 0.79$ & $-108.30 \pm 0.93$ & $3.50 \pm 0.29$ & $-37.5 \pm 1.1$ \\
 & \multicolumn{2}{c}{$\theta=3.07552 \pm 0.00058$\,\arcsec} & \multicolumn{2}{c}{$s=211.6 \pm 3.0$\,au} & \multicolumn{2}{c}{$v_\textrm{rel}=3.7 \pm 1.2$\,km\,s$^{-1}$}\\
\hline 
5436014972680358272 & $38.082 \pm 0.030$ & $26.259 \pm 0.021$ & $-339.427 \pm 0.036$ & $181.405 \pm 0.046$ & $-42.252 \pm 0.034$ & $22.581 \pm 0.019$ \\
5436014972680358784 & $38.060 \pm 0.031$ & $26.275 \pm 0.021$ & $-328.220 \pm 0.037$ & $167.831 \pm 0.048$ & $-40.881 \pm 0.033$ & $20.904 \pm 0.018$ \\
 & \multicolumn{2}{c}{$\theta=4.427489 \pm 0.000033$\,\arcsec} & \multicolumn{2}{c}{$s=116.262 \pm 0.093$\,au} & \multicolumn{2}{c}{$v_\textrm{rel}=2.166 \pm 0.050$\,km\,s$^{-1}$}\\
\hline 
3970693313784409344 & $9.74 \pm 0.30$ & $102.6 \pm 3.2$ & $-62.06 \pm 0.49$ & $2.09 \pm 0.36$ & $-30.19 \pm 0.97$ & $1.02 \pm 0.18$ \\
3970693313782310656 & $9.92 \pm 0.26$ & $100.8 \pm 2.6$ & $-60.16 \pm 0.42$ & $2.62 \pm 0.31$ & $-28.76 \pm 0.77$ & $1.25 \pm 0.15$ \\
 & \multicolumn{2}{c}{$\theta=2.68969 \pm 0.00029$\,\arcsec} & \multicolumn{2}{c}{$s=276.1 \pm 8.6$\,au} & \multicolumn{2}{c}{$v_\textrm{rel}=1.5 \pm 1.2$\,km\,s$^{-1}$}\\
\hline 
5874024774146311808 & $8.69 \pm 0.11$ & $115.0 \pm 1.5$ & $-33.69 \pm 0.15$ & $-53.24 \pm 0.18$ & $-18.37 \pm 0.25$ & $-29.03 \pm 0.39$ \\
5874024769842933760 & $8.424 \pm 0.080$ & $118.7 \pm 1.1$ & $-33.808 \pm 0.099$ & $-54.57 \pm 0.11$ & $-19.02 \pm 0.19$ & $-30.71 \pm 0.30$ \\
 & \multicolumn{2}{c}{$\theta=2.569850 \pm 0.000056$\,\arcsec} & \multicolumn{2}{c}{$s=295.6 \pm 3.9$\,au} & \multicolumn{2}{c}{$v_\textrm{rel}=1.80 \pm 0.56$\,km\,s$^{-1}$}\\
\hline 
5886778250054825344 & $9.40 \pm 0.22$ & $106.4 \pm 2.5$ & $-32.52 \pm 0.46$ & $-61.25 \pm 0.36$ & $-16.40 \pm 0.45$ & $-30.89 \pm 0.74$ \\
5886777494107570176 & $8.98 \pm 0.26$ & $111.3 \pm 3.2$ & $-32.25 \pm 0.55$ & $-59.38 \pm 0.43$ & $-17.01 \pm 0.57$ & $-31.33 \pm 0.94$ \\
 & \multicolumn{2}{c}{$\theta=2.78136 \pm 0.00015$\,\arcsec} & \multicolumn{2}{c}{$s=295.8 \pm 6.9$\,au} & \multicolumn{2}{c}{$v_\textrm{rel}=0.8 \pm 1.2$\,km\,s$^{-1}$}\\
\hline 
4450425359563998720 & $11.69 \pm 0.13$ & $85.57 \pm 0.94$ & $-19.16 \pm 0.18$ & $50.88 \pm 0.11$ & $-7.77 \pm 0.11$ & $20.64 \pm 0.23$ \\
4450425359563998848 & $11.58 \pm 0.15$ & $86.4 \pm 1.1$ & $-18.26 \pm 0.21$ & $48.39 \pm 0.13$ & $-7.47 \pm 0.13$ & $19.81 \pm 0.26$ \\
 & \multicolumn{2}{c}{$\theta=2.61608 \pm 0.00011$\,\arcsec} & \multicolumn{2}{c}{$s=223.9 \pm 2.5$\,au} & \multicolumn{2}{c}{$v_\textrm{rel}=0.88 \pm 0.37$\,km\,s$^{-1}$}\\
\hline 
1408135749896104192 & $25.377 \pm 0.027$ & $39.406 \pm 0.042$ & $22.799 \pm 0.048$ & $-70.057 \pm 0.051$ & $4.259 \pm 0.010$ & $-13.087 \pm 0.017$ \\
1408135749896103936 & $25.342 \pm 0.029$ & $39.460 \pm 0.044$ & $15.026 \pm 0.051$ & $-70.573 \pm 0.057$ & $2.811 \pm 0.010$ & $-13.201 \pm 0.018$ \\
 & \multicolumn{2}{c}{$\theta=5.611901 \pm 0.000025$\,\arcsec} & \multicolumn{2}{c}{$s=221.14 \pm 0.23$\,au} & \multicolumn{2}{c}{$v_\textrm{rel}=1.453 \pm 0.014$\,km\,s$^{-1}$}\\
\hline 

\end{tabular}
\end{center}
\end{table*}

\begin{table*}
\contcaption{}
\begin{center}
\begin{tabular}{c c c c c c c c}
\hline
GaiaID & $\varpi$ (mas) & $d$ (pc) & $\mu_\textrm{RA}$ (mas\,yr$^{-1}$) & $\mu_\textrm{Dec}$ (mas\,yr$^{-1}$) & $v_\textrm{RA}$ (km\,s$^{-1}$) & $v_\textrm{Dec}$ (km\,s$^{-1}$)\\
\hline
2269560710341753728 & $17.667 \pm 0.086$ & $56.60 \pm 0.28$ & $-43.88 \pm 0.15$ & $-135.61 \pm 0.20$ & $-11.774 \pm 0.070$ & $-36.39 \pm 0.18$ \\
2269560710340167296 & $17.77 \pm 0.16$ & $56.26 \pm 0.50$ & $-42.37 \pm 0.28$ & $-137.58 \pm 0.37$ & $-11.30 \pm 0.13$ & $-36.69 \pm 0.34$ \\
 & \multicolumn{2}{c}{$\theta=3.22349 \pm 0.00017$\,\arcsec} & \multicolumn{2}{c}{$s=182.46 \pm 0.89$\,au} & \multicolumn{2}{c}{$v_\textrm{rel}=0.56 \pm 0.14$\,km\,s$^{-1}$}\\
\hline 
2150591795574568192 & $20.93 \pm 0.19$ & $47.78 \pm 0.43$ & $-156.56 \pm 0.42$ & $-160.73 \pm 0.39$ & $-35.46 \pm 0.33$ & $-36.40 \pm 0.34$ \\
2150591799870095872 & $20.71 \pm 0.17$ & $48.29 \pm 0.39$ & $-155.79 \pm 0.35$ & $-158.73 \pm 0.33$ & $-35.66 \pm 0.30$ & $-36.34 \pm 0.31$ \\
 & \multicolumn{2}{c}{$\theta=5.51585 \pm 0.00014$\,\arcsec} & \multicolumn{2}{c}{$s=263.5 \pm 2.4$\,au} & \multicolumn{2}{c}{$v_\textrm{rel}=0.22 \pm 0.29$\,km\,s$^{-1}$}\\
\hline 
4257063458004688896 & $25.24 \pm 0.10$ & $39.63 \pm 0.16$ & $118.65 \pm 0.20$ & $115.82 \pm 0.19$ & $22.287 \pm 0.097$ & $21.756 \pm 0.095$ \\
4257063453704172416 & $25.131 \pm 0.080$ & $39.79 \pm 0.13$ & $117.04 \pm 0.17$ & $122.90 \pm 0.18$ & $22.077 \pm 0.077$ & $23.181 \pm 0.081$ \\
 & \multicolumn{2}{c}{$\theta=2.74113 \pm 0.00011$\,\arcsec} & \multicolumn{2}{c}{$s=108.62 \pm 0.44$\,au} & \multicolumn{2}{c}{$v_\textrm{rel}=1.44 \pm 0.11$\,km\,s$^{-1}$}\\
\hline 
6650228409376970880 & $16.34 \pm 0.30$ & $61.2 \pm 1.1$ & $-90.47 \pm 0.30$ & $-29.11 \pm 0.29$ & $-26.25 \pm 0.50$ & $-8.45 \pm 0.18$ \\
6650228409380791040 & $16.53 \pm 0.12$ & $60.49 \pm 0.42$ & $-87.42 \pm 0.12$ & $-23.54 \pm 0.12$ & $-25.07 \pm 0.18$ & $-6.750 \pm 0.058$ \\
 & \multicolumn{2}{c}{$\theta=3.66520 \pm 0.00012$\,\arcsec} & \multicolumn{2}{c}{$s=224.3 \pm 4.2$\,au} & \multicolumn{2}{c}{$v_\textrm{rel}=2.07 \pm 0.44$\,km\,s$^{-1}$}\\
\hline 
2080526555267049984 & $34.718 \pm 0.072$ & $28.803 \pm 0.060$ & $-455.33 \pm 0.14$ & $-401.45 \pm 0.15$ & $-62.17 \pm 0.13$ & $-54.81 \pm 0.12$ \\
2080526555267050496 & $34.731 \pm 0.061$ & $28.793 \pm 0.051$ & $-453.28 \pm 0.12$ & $-404.61 \pm 0.12$ & $-61.87 \pm 0.11$ & $-55.225 \pm 0.099$ \\
 & \multicolumn{2}{c}{$\theta=8.637912 \pm 0.000088$\,\arcsec} & \multicolumn{2}{c}{$s=248.80 \pm 0.52$\,au} & \multicolumn{2}{c}{$v_\textrm{rel}=0.510 \pm 0.033$\,km\,s$^{-1}$}\\
\hline 
6466132607691464320 & $20.20 \pm 0.23$ & $49.50 \pm 0.57$ & $12.38 \pm 0.28$ & $-59.52 \pm 0.39$ & $2.904 \pm 0.074$ & $-13.97 \pm 0.18$ \\
6466132607693274496 & $20.06 \pm 0.16$ & $49.85 \pm 0.39$ & $4.24 \pm 0.18$ & $-54.13 \pm 0.20$ & $1.002 \pm 0.042$ & $-12.79 \pm 0.11$ \\
 & \multicolumn{2}{c}{$\theta=2.11378 \pm 0.00013$\,\arcsec} & \multicolumn{2}{c}{$s=104.6 \pm 1.2$\,au} & \multicolumn{2}{c}{$v_\textrm{rel}=2.24 \pm 0.15$\,km\,s$^{-1}$}\\
\hline 
1801239293158983168 & $15.02 \pm 0.14$ & $66.58 \pm 0.64$ & $-58.22 \pm 0.27$ & $-103.72 \pm 0.33$ & $-18.37 \pm 0.19$ & $-32.73 \pm 0.33$ \\
1801239297453428992 & $14.79 \pm 0.11$ & $67.63 \pm 0.53$ & $-60.42 \pm 0.22$ & $-105.84 \pm 0.26$ & $-19.37 \pm 0.17$ & $-33.93 \pm 0.28$ \\
 & \multicolumn{2}{c}{$\theta=3.16333 \pm 0.00013$\,\arcsec} & \multicolumn{2}{c}{$s=210.6 \pm 2.0$\,au} & \multicolumn{2}{c}{$v_\textrm{rel}=1.56 \pm 0.48$\,km\,s$^{-1}$}\\
\hline 
2841680934336159104 & $21.76 \pm 0.24$ & $45.96 \pm 0.50$ & $-85.16 \pm 0.32$ & $-179.28 \pm 0.22$ & $-18.56 \pm 0.21$ & $-39.06 \pm 0.42$ \\
2841680934336158976 & $21.70 \pm 0.14$ & $46.09 \pm 0.30$ & $-72.14 \pm 0.22$ & $-177.52 \pm 0.14$ & $-15.76 \pm 0.11$ & $-38.79 \pm 0.25$ \\
 & \multicolumn{2}{c}{$\theta=3.71565 \pm 0.00013$\,\arcsec} & \multicolumn{2}{c}{$s=170.8 \pm 1.8$\,au} & \multicolumn{2}{c}{$v_\textrm{rel}=2.81 \pm 0.28$\,km\,s$^{-1}$}\\
\hline 
1997383581920045312 & $19.96 \pm 0.16$ & $50.09 \pm 0.41$ & $128.12 \pm 0.22$ & $62.29 \pm 0.18$ & $30.42 \pm 0.26$ & $14.79 \pm 0.13$ \\
1997383581913649152 & $19.95 \pm 0.15$ & $50.12 \pm 0.37$ & $135.64 \pm 0.20$ & $68.36 \pm 0.16$ & $32.23 \pm 0.24$ & $16.24 \pm 0.13$ \\
 & \multicolumn{2}{c}{$\theta=2.41721 \pm 0.00010$\,\arcsec} & \multicolumn{2}{c}{$s=121.1 \pm 1.0$\,au} & \multicolumn{2}{c}{$v_\textrm{rel}=2.31 \pm 0.38$\,km\,s$^{-1}$}\\
\hline 
2017484922210746368 & $17.68 \pm 0.26$ & $56.57 \pm 0.83$ & $346.07 \pm 0.49$ & $211.20 \pm 0.36$ & $92.8 \pm 1.4$ & $56.63 \pm 0.84$ \\
2017484922214792576 & $17.33 \pm 0.12$ & $57.69 \pm 0.40$ & $344.68 \pm 0.20$ & $209.45 \pm 0.17$ & $94.26 \pm 0.65$ & $57.28 \pm 0.40$ \\
 & \multicolumn{2}{c}{$\theta=2.63999 \pm 0.00020$\,\arcsec} & \multicolumn{2}{c}{$s=149.3 \pm 2.2$\,au} & \multicolumn{2}{c}{$v_\textrm{rel}=1.6 \pm 1.8$\,km\,s$^{-1}$}\\

\hline
\end{tabular}
\end{center}
\end{table*}

Next, we search for tertiary companions around the inner double WDs, by querying the entire \textit{Gaia} DR2 database. Assuming a uniform distribution of orbital eccentricities, as supported by observations \citep{Duchene_2013, Tokovinin_2016}, the mean eccentricity is $\left<e\right>=1/2$, and the apocentre separation is therefore, on average, $\left(1+e\right)/\left(1-e\right)=3$ times the pericentre separation. The apocentre separation of the tertiary star in the collisional model can thus be as high as $\sim 30a$, and we therefore search for tertiaries projected within 9000\,au of the inner, $a<300$\,au, double WDs. 

For the tertiaries we include sources only if they satisfy $\texttt{parallax\_over\_error} > 5$, Eq.~\ref{eq:astrometric} (astrometric artefacts removal), and Eq.~\ref{eq:distance} (distance from the first WD in the inner binary), but this time we allow only a $2 \sigma_{\Delta d}$ difference in the distances. None of the 27 resolved inner WD binaries has a candidate tertiary companion projected within 9000\,au. With Poisson statistics, this null detection implies, at 95~per cent confidence, the existence of $<3$ tertiaries, or an upper limit of $<11$~per cent on the fraction of $a<300$\,au-separation double WDs with a tertiary companion with $r_\textrm{peri,out}=\left(3-10\right)a$.

\subsection{\textit{Gaia} tertiaries to SPY double WDs}
To further explore the potential triple SN Ia progenitor landscape, we have searched for tertiaries around a second sample of close double-WD binaries, this time unresolved WD binary candidates from the Supernova Progenitor surveY \citep[SPY;][]{Napiwotzki_2001}. The SPY survey was a few-epoch spectroscopic survey of $\sim$800 bright ($V\sim 16$\,mag) WDs conducted with the European Southern Observatory 8\,m Very Large Telescope, with the objective of using radial-velocity differences between epochs to identify close double-WD systems that will merge within a Hubble time. In \citet{Maoz_2017} we measured the maximal changes in radial velocity (\drvm) between epochs, and modelled the observed \drvm\ statistics via Monte Carlo simulations \citep{Maoz_2012, Badenes_2012}, to constrain the population characteristics of double WDs. We identified 43 high-\drvm\ systems as likely double-WD candidates \citep[see table 1 of][]{Maoz_2017}, and an additional double-lined system that was not included in the candidate list because of its low \drvm\ (HE\,0315$-$0118).

Since the \textit{Gaia} DR2 astrometric model does not take binarity and potential astrometric wobble into account, it might fail, or not fail, but report spurious results in case of unresolved binaries \citep[e.g.][]{Hollands_2018}. We have verified that full five-parameter astrometric solutions are reported for all 44 SPY systems. To further check for spurious distances, we compared the astrometric distances derived from \textit{Gaia}'s parallaxes with the photometric distances for these WDs. We used the spectroscopic effective temperatures and surface gravities derived by \citet{Koester_2009} to generate synthetic WD model spectra using the spectral synthesis program \textsc{Synspec} \citep[version 50;][]{Hubeny_2011}, based on model atmospheres created by the \textsc{Tlusty} program \citep[version 205;][]{Hubeny_1988, Hubeny_1995, Hubeny_2017a}. The luminosities were then estimated by assuming a mass-radius relation \citep[`thick' hydrogen-dominated carbon-oxygen WD cooling tracks\footnote{\href{http://www.astro.umontreal.ca/bergeron/CoolingModels/}{http://www.astro.umontreal.ca/bergeron/CoolingModels/}} from][]{Fontaine_2001}. Finally, the photometric distances were calculated using the $V$/$B$-band magnitudes reported in \citet{Koester_2009}. We find that the photometric distances of all of the SPY WDs are consistent with their \textit{Gaia} distances, with a scatter $\sim 10$~per cent, and with few or no obvious outliers that could be indicative of spurious \textit{Gaia} distances. Nonetheless, future \textit{Gaia} data releases that will include binarity in the astrometric model are required in order to verify that none of the WD distances are erroneous due to unresolved binarity.

As with the previous sample, we have searched with \textit{Gaia} for tertiaries within 3000\,au of these 44 candidate close-double-WDs from SPY. Contrary to the previous sample, the triple systems turned up here are not actual candidate progenitors for the collisional-triple scenario, since their inner separations are of order $0.1-4$\,au, whereas the tertiary separation is of order $10^{3-4}$ larger, rather than just $3-30$ times larger, as required by the model. The tertiaries within only 120\,au, which would be have been relevant to the model, are well below the \textit{Gaia} angular resolution limit at the typical distances of the SPY WDs ($\sim 10-250$\,pc). Nevertheless, it is instructive to estimate the frequency of these ``extreme'' (rather than ``mild'') hierarchical triples, to learn about the triple population in this different hierarchy range.

\begin{figure}
\includegraphics[width=\columnwidth]{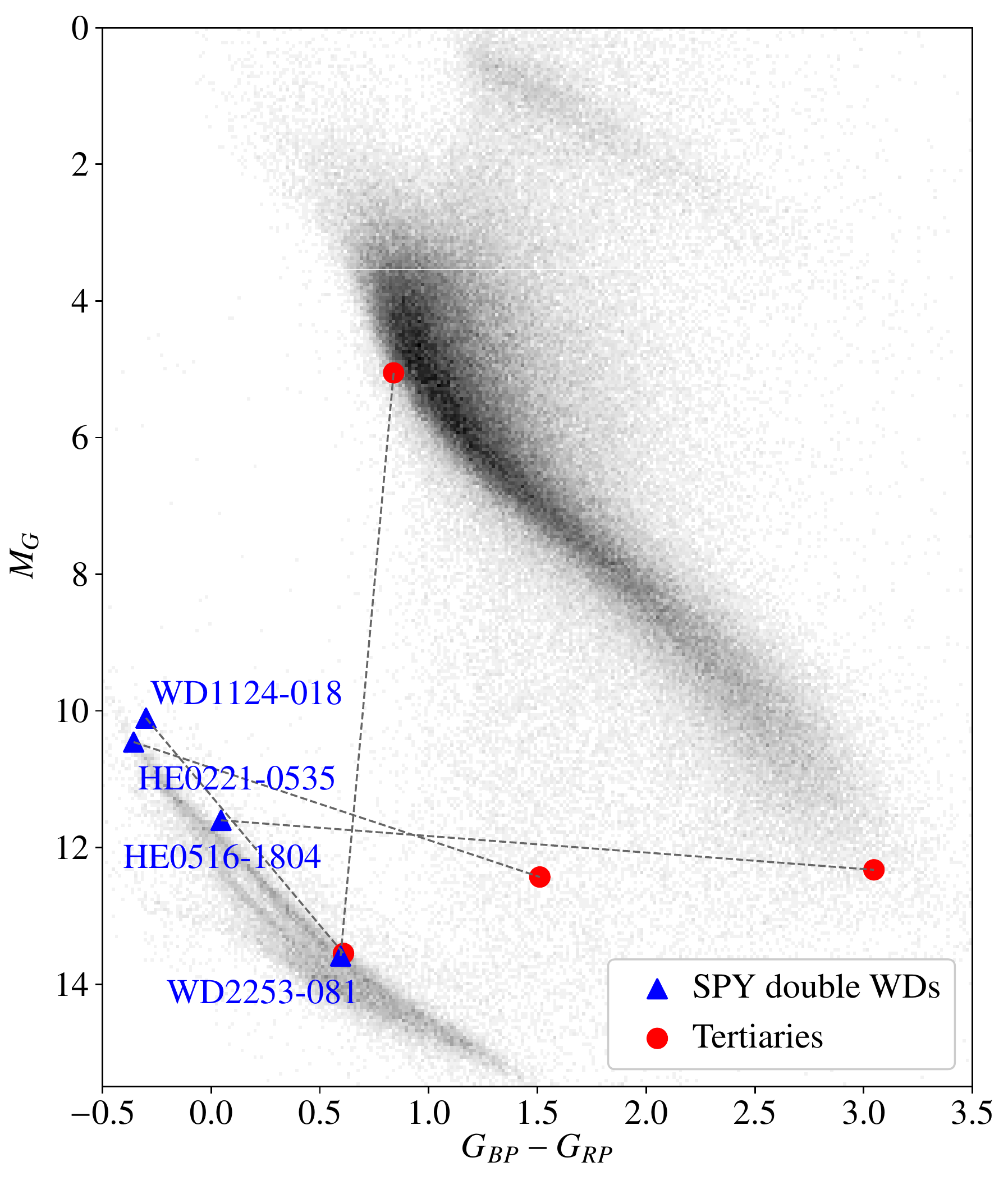}
\caption{Candidate unresolved double WDs from the SPY sample (blue triangles) with a tertiary stellar companion in \textit{Gaia} with projected separations $s<3000$\,au (red circles). Pairs are connected by dashed lines. The number density of stars from a random \textit{Gaia} sample and the full 120\,pc WD sample is shown in grayscale.}
\label{fig:SPY}
\end{figure}

Among the 44 SPY double-WD candidates, we find four with tertiaries projected within $<3000$\,au separation, satisfying $\texttt{parallax\_over\_error} > 5$, Eq.~\ref{eq:astrometric} (astrometric artefacts removal), and Eq.~\ref{eq:distance} (distance from the first WD in the inner binary, within $2 \sigma_{\Delta d}$). We list them in Table~\ref{tab:SPY}. Their location on the \textit{Gaia} colour-magnitude diagram is shown in Fig.~\ref{fig:SPY}. At 3000\,au separation, the maximum velocity difference due to orbital motion is $\sim 1.5$\,km\,s$^{-1}$, and again all four candidate tertiaries are consistent with a velocity difference that is lower than this, arguing that these are true bound systems. The M-dwarf tertiary we find for HE\,0516$-$1804 could be the source of the infrared excess we found in its photometry in \citet{Maoz_2017}. We note that the known M-dwarf tertiary candidate of WD\,0326$-$273 \citep[][and references therein]{Nelemans_2005} failed to pass our consistent parallax criterion (Eq.~\ref{eq:distance}), with $\sigma_{\Delta d} \sim 5$.

This result suggests that $\lesssim 9$~per cent of separation-$a\sim 0.1-4$\,au double WDs have tertiaries within separations of 3000\,au. Again, the population of tertiaries considered around this sample is much more extensive than those with a role in the collisional-triple SN Ia model. 

\begin{table*}
\caption{Candidate unresolved double-WD systems from the SPY sample having a \textit{Gaia} tertiary companion with projected separation $s<3000$\,au. Each triple system appears as a pair of rows separated by a line. The first line of each pair represents the inner (unresolved) double WD, and the second line describes the tertiary star. Columns as in Table~\ref{tab:DWD}.}
\label{tab:SPY}
\begin{center}
\begin{tabular}{c c c c c c c c c}
\hline
Name & GaiaID & $\varpi$ (mas) & $d$ (pc) & $\mu_\textrm{RA}$ (mas\,yr$^{-1}$) & $\mu_\textrm{Dec}$ (mas\,yr$^{-1}$) & $v_\textrm{RA}$ (km\,s$^{-1}$) & $v_\textrm{Dec}$ (km\,s$^{-1}$)\\
\hline
HE0221-0535 & 2488754366292015744 & $8.97 \pm 0.11$ & $111.5 \pm 1.3$ & $-17.85 \pm 0.13$ & $5.69 \pm 0.12$ & $-9.43 \pm 0.13$ & $3.006 \pm 0.075$ \\
 & 2488754366292015872 & $8.40 \pm 0.29$ & $119.0 \pm 4.1$ & $-18.72 \pm 0.33$ & $5.97 \pm 0.28$ & $-10.56 \pm 0.41$ & $3.37 \pm 0.20$ \\
 & & \multicolumn{2}{c}{$\theta=2.68970 \pm 0.00019$\,\arcsec} & \multicolumn{2}{c}{$s=299.8 \pm 3.6$\,au} & \multicolumn{2}{c}{$v_\textrm{rel}=1.19 \pm 0.45$\,km\,s$^{-1}$}\\
\hline 
HE0516-1804 & 2981590730954538112 & $11.884 \pm 0.061$ & $84.15 \pm 0.43$ & $9.90 \pm 0.10$ & $-30.24 \pm 0.11$ & $3.951 \pm 0.046$ & $-12.061 \pm 0.077$ \\
 & 2981590730954537984 & $11.679 \pm 0.099$ & $85.63 \pm 0.72$ & $11.17 \pm 0.16$ & $-31.72 \pm 0.18$ & $4.535 \pm 0.076$ & $-12.87 \pm 0.13$ \\
 & & \multicolumn{2}{c}{$\theta=2.289622 \pm 0.000094$\,\arcsec} & \multicolumn{2}{c}{$s=192.66 \pm 0.99$\,au} & \multicolumn{2}{c}{$v_\textrm{rel}=1.00 \pm 0.15$\,km\,s$^{-1}$}\\
\hline 
WD1124-018 & 3796545519645331584 & $5.47 \pm 0.11$ & $182.8 \pm 3.8$ & $-35.87 \pm 0.18$ & $11.95 \pm 0.12$ & $-31.08 \pm 0.66$ & $10.36 \pm 0.24$ \\
 & 3796545515349746432 & $5.84 \pm 0.70$ & $171 \pm 21$ & $-35.66 \pm 0.98$ & $11.64 \pm 0.63$ & $-29.0 \pm 3.6$ & $9.5 \pm 1.2$ \\
 & & \multicolumn{2}{c}{$\theta=9.01849 \pm 0.00056$\,\arcsec} & \multicolumn{2}{c}{$s=1649 \pm 34$\,au} & \multicolumn{2}{c}{$v_\textrm{rel}=2.3 \pm 3.8$\,km\,s$^{-1}$}\\
\hline 
WD2253-081 & 2610488514148351360 & $27.78 \pm 0.11$ & $35.99 \pm 0.14$ & $549.25 \pm 0.13$ & $-46.94 \pm 0.12$ & $93.71 \pm 0.36$ & $-8.009 \pm 0.037$ \\
 & 2610488514148352000 & $27.732 \pm 0.075$ & $36.060 \pm 0.097$ & $551.706 \pm 0.090$ & $-48.924 \pm 0.084$ & $94.31 \pm 0.25$ & $-8.363 \pm 0.027$ \\
 & & \multicolumn{2}{c}{$\theta=41.656756 \pm 0.000075$\,\arcsec} & \multicolumn{2}{c}{$s=1499.3 \pm 5.7$\,au} & \multicolumn{2}{c}{$v_\textrm{rel}=0.69 \pm 0.40$\,km\,s$^{-1}$}\\

\hline
\end{tabular}
\end{center}
\end{table*}

\section{Discussion}   
We have used \textit{Gaia} to search for candidate triple stellar systems similar to those envisaged in the collisional triple SN Ia progenitor model, and to assess their abundance. \textit{Gaia}, due to its angular resolution limits, is strongly biased against detection of the separation $a<300$\,au inner WD binaries of the proposed triple systems. We have therefore focused on two samples of ``known'' double WDs, and used \textit{Gaia} to search for a putative wide tertiary member out to 9000\,au separations, where \textit{Gaia} completeness is high. Our first inner-double-WD sample consists of 27 WD pairs with projected separations $s<300$\,au, found in the \textit{Gaia} DR2 database among $\sim 17,400$ WDs within a 120\,pc distance. These 27 systems are the small fraction of the actual number of double WDs having such separations that are detected as binaries, because of \textit{Gaia}'s low sensitivity in this regime. 

Around this sample of 27 \textit{Gaia}-selected inner-WD-binaries, we find no tertiary stars of any kind (WD or main-sequence, down to the lowest stellar masses) out to 9000\,au projected separations. Such systems, had they been found, could correspond closely in separation and hierarchy to the SN Ia progenitors in the model. Their non-detection sets a 95~per-cent-confidence upper limit of 11~per cent on the fraction of $a<300$\,au WD binaries that are orbited by a tertiary with pericentre separation $3-10$ times the inner-binary separation. \citet{Maoz_2018} have combined  the \drvm\ results of \citet{Badenes_2012} and \citet{Maoz_2017}, to deduce that the fraction of WDs in binaries in the range 0.01 to 4\,au is $10\pm2.5$~per cent, with a separation distribution in this range consistent with a power law, $a^\alpha$, with $\alpha=-1.3\pm0.2$. \citet{ElBadry_2018}, analysing the resolved binary WD population that they find with \textit{Gaia}, and after accounting for selection and incompleteness effects, find a power-law separation distribution with a slope, again, close to $\alpha=-1$ (i.e. equal numbers per logarithmic interval in separation) in the range $a=50-1500$\,au. Considering then that there is a similar number of decades in separation between $0.01-4$\,au \citep{Maoz_2018} and between $1-300$\,au (the range of inner-binary separations in the \citet{Katz_2012} model), implies that $\sim 10$~per cent of all WDs are in binaries with $a=1-300$\,au. Combined with our present result that, at most, 11~per cent of such binaries host a tertiary companion out to 9000\,au, we conclude that the fraction among WDs of triple systems with the architecture required by the collisional-triple model is $<1$~per cent, at the 95~per cent confidence level. This triple fraction is at least a factor $\sim 30-55$ times lower than the fraction required by the model ($\sim 30-55$~per cent, see Section~\ref{sec:Intro}). 

A second double-WD sample that we have studied consists of 44 unresolved, separation $a\sim 0.1-4$\,au candidate WD binaries that have emerged, based on radial-velocity variations, from SPY. We searched the SPY candidate WD binaries for tertiary companions, this time out to 3000\,au separations, i.e., a much larger range than relevant for the collisional model. Here, we found four candidate tertiary companions (two main-sequence stars, a WD, and an uncertain case, perhaps a subdwarf) with actual projected separations from 190 to 1650\,au. Considering the WD binarity fraction of roughly 4~per cent for every decade of separation (see above), the 9~per cent fraction of tertiaries (of all kinds) in this separation range around short-orbit WD binaries is not surprising. Combined with the abundance of inner-WD binaries, we again get a 1~per cent upper limit on the fraction of WDs that are in triples, consisting of an inner double WD plus a tertiary (although now triples of a different hierarchy than those required by the collisional-triple model).

We conclude that there is at least a 1.5 order of magnitude deficit, and likely more, in the fraction of WDs that are in the types of triple systems required by the collisional-triple model. The double WDs in our first, $s<300$\,au, sample have actual separations in the $100-300$\,au range. A remaining ``hiding place'' for the progenitor systems of the model could be in the unexplored parameter range of $a\sim 1-100$\,au WD binaries. A large fraction of all WDs would need to be in binaries with this separation range, \textit{and} a large fraction of those same binaries would need to have tertiaries at separations $(3-10)a$. The first condition is unlikely to be true, given the similar separation distributions (about flat in log separation, see above) followed by binary WDs within the $a=0.01-4$\,au range and the $a=50-1500$\,au range. The second condition is also unlikely, given the rarity of tertiary stars around WD binaries with those separation ranges, as shown here. A different progenitor and explosion scenario is most probably at the root of SN Ia explosions.

\section*{Acknowledgements}
We thank Boaz Katz, Julio Chanam\'e, J. J. Hermes, and the anonymous referee for valuable comments. This work was supported by a grant from the Israel Science Foundation (ISF). This work has made use of data from the European Space Agency (ESA) mission {\it Gaia} (\href{https://www.cosmos.esa.int/gaia}{https://www.cosmos.esa.int/gaia}), processed by the \textit{Gaia} Data Processing and Analysis Consortium (DPAC, \href{https://www.cosmos.esa.int/web/gaia/dpac/consortium}{https://www.cosmos.esa.int/web/gaia/dpac/consortium}). Funding for the DPAC has been provided by national institutions, in particular the institutions participating in the {\it Gaia} Multilateral Agreement.

This research made use of \textsc{astropy}\footnote{\href{http://www.astropy.org}{http://www.astropy.org}}, a community-developed core \textsc{python} package for Astronomy \citep{Astropy_2013, Astropy_2018}, \textsc{astroquery} \citep{Astroquery_2013}, \textsc{matplotlib} \citep{Hunter_2007}, \textsc{numpy} \citep{Numpy_2006, Numpy_2011}, \textsc{uncertainties}\footnote{\href{http://pythonhosted.org/uncertainties/}{http://pythonhosted.org/uncertainties/}}, a \textsc{python} package for calculations with uncertainties by Eric O. Lebigot, and \textsc{topcat} \citep{Taylor_2005}, a tool for operations on catalogues and tables.



\bibliographystyle{mnras}
\bibliography{gaiatriples} 




%
%


\bsp	
\label{lastpage}
\end{document}